\documentclass[aps,letterpaper,twocolumn,prb,footinbib,floatfix,showpacs,superscriptaddress]{revtex4-1}
\usepackage{amsmath}
\usepackage{amsfonts}
\usepackage{amssymb}
\usepackage[scanall]{psfrag}
\usepackage{psfrag}
\usepackage{graphicx}
\usepackage{color}
\usepackage[T1]{fontenc}
\usepackage{times}
\usepackage[colorlinks,bookmarks=true,citecolor=blue,linkcolor=red,urlcolor=blue]{hyperref}

\begin{document}
\newcommand{\of}[1]{\left( #1 \right)}
\newcommand{\sqof}[1]{\left[ #1 \right]}
\newcommand{\abs}[1]{\left| #1 \right|}
\newcommand{\avg}[1]{\left< #1 \right>}
\newcommand{\cuof}[1]{\left \{ #1 \right \} }
\newcommand{\bra}[1]{\left < #1 \right | }
\newcommand{\ket}[1]{\left | #1 \right > }
\newcommand{\pil}{\frac{\pi}{L}}
\newcommand{\bx}{\mathbf{x}}
\newcommand{\by}{\mathbf{y}}
\newcommand{\bk}{\mathbf{k}}
\newcommand{\bp}{\mathbf{p}}
\newcommand{\bl}{\mathbf{l}}
\newcommand{\bq}{\mathbf{q}}
\newcommand{\bs}{\mathbf{s}}
\newcommand{\psibar}{\overline{\psi}}
\newcommand{\svec}{\overrightarrow{\sigma}}
\newcommand{\dvec}{\overrightarrow{\partial}}
\newcommand{\bA}{\mathbf{A}}
\newcommand{\bdelta}{\mathbf{\delta}}
\newcommand{\bK}{\mathbf{K}}
\newcommand{\bQ}{\mathbf{Q}}
\newcommand{\bG}{\mathbf{G}}
\newcommand{\bw}{\mathbf{w}}
\newcommand{\bL}{\mathbf{L}}
\newcommand{\ohat}{\widehat{O}}
\newcommand{\up}{\uparrow}
\newcommand{\down}{\downarrow}
\newcommand{\MM}{\mathcal{M}}
\author{Zhao Liu}
\affiliation{Department of Electrical Engineering, Princeton University, Princeton, New Jersey 08544, USA}
\email{zhaol@princeton.edu}
\affiliation{Beijing Computational Science Research Center, Beijing, 100084, China}
\author{Emil J. Bergholtz}
\affiliation{Dahlem Center for Complex Quantum Systems and Institut f\"ur Theoretische Physik, Freie Universit\"at Berlin, Arnimallee 14, 14195 Berlin, Germany}
\email{ejb@physik.fu-berlin.de}
\author{Eliot Kapit}
\affiliation{Rudolf Peierls Center for Theoretical Physics, Oxford University, 1 Keble Road, Oxford OX1 3NP, United Kingdom}
\email{eliot.kapit@physics.ox.ac.uk}

\title{Non-Abelian fractional Chern insulators from long-range interactions}

\begin{abstract}

The recent theoretical discovery of fractional Chern insulators (FCIs) has provided an important new way to realize topologically ordered states in lattice models. In earlier works, on-site and nearest neighbor Hubbard-like interactions have been used extensively to stabilize Abelian FCIs in systems with nearly flat, topologically nontrivial bands. However, attempts to use two-body interactions to stabilize non-Abelian FCIs, where the ground state in the presence of impurities can be massively degenerate and manipulated through anyon braiding, have proven very difficult in uniform lattice systems. Here, we study the remarkable effect of long-range interactions in a lattice model that possesses an exactly flat lowest band with a unit Chern number. When spinless bosons with two-body long-range interactions partially fill the lowest Chern band, we find convincing evidence of gapped, bosonic Read-Rezayi (RR) phases with non-Abelian anyon statistics. We characterize these states through studying topological degeneracies, the overlap between the ground states of two-body interactions and the exact RR ground states of three- and four-body interactions, and state counting in the particle-cut entanglement spectrum. Moreover, we demonstrate how an approximate lattice form of Haldane's pseudopotentials, analogous to that in the continuum, can be used as an efficient guiding principle in the search for lattice models with stable non-Abelian phases.

\end{abstract}
\maketitle

\section{Introduction}

The possibility of realizing lattice generalizations of fractional quantum Hall states, topologically ordered states where the fundamental collective excitations of the system are anyons with fractional statistics, has been an exciting recent development.\cite{kapitmueller,sungu,tangmei,neupertprl,rbprx,qi1,lauchli,ifwprl,rbwannier,liuhighC,wubloch,yaodipo,cooperof} These states, known more generally as fractional Chern insulators (FCIs), are highly exotic. For instance, a bosonic FCI, which could be realized in cold atoms or qubit arrays, is a highly correlated fractional quantum liquid of bosons in a strong (artificial) magnetic field. In the condensed matter context, similar phenomenology, albeit with fermions as basic constituents, may arise from a combination of ferromagnetism and spin-orbit coupling. In both settings, the FCIs are stabilized by interactions on the lattice scale, which, in materials, is typically three orders of magnitude larger than the cyclotron frequency in two-dimensional (2D) electron gas experiments. Thus, the strong correlations of an electronic fermionic FCI could potentially persist at temperatures far higher than the subkelvin limits of the 2D electron gas in strong magnetic fields.

Our understanding of FCIs is predominantly guided by the analogy to continuum Landau levels, where the elegant symmetries and mathematical features of the system, such as the exact band flatness and analytic structure of the wave functions in the symmetric gauge, open the door to many exact and nonperturbative results that would be inaccessible in more general cases. In the low-flux limit, the effect of the lattice is a small perturbation to the continuum problem,\cite{sorensen,hafezi} but at higher flux densities topological flat bands and continuum Landau levels differ significantly in details, despite their topological equivalence. For instance, translation and rotation are broken into discrete symmetries within the flat bands, and the Berry curvature is necessarily inhomogeneous in reciprocal space in any lattice model. Despite these complications and new, lattice-specific competing instabilities, intense recent theoretical research on topological flat band models has demonstrated that a number of FCI counterparts of fractional quantum Hall states emerge at low energies (see Refs. \onlinecite{fcireview1,fcireview2} for reviews). Laughlin states, which have Abelian anyon excitations, are particularly robust and can be realized in both bosons and fermions for many different lattices and flux densities.

However, a striking limitation to this beautiful analogy is that it has proven very difficult to actually construct realistic models that stabilize lattice versions of non-Abelian FCIs, where the fluid's anyon excitations carry additional degeneracies which can be manipulated through adiabatically exchanging and braiding the anyons. In fact, non-Abelian phases in fractional Chern bands have only been firmly established in the presence of dominant multiparticle interactions \cite{wangyao,wubernevig,bernevigregnault,liuadia,kapitsimon,hafeziadhikari} that add an additional energy cost for three or more particles to occupy the same (or neighboring) sites, and analogous results have been obtained in chiral spin liquid models with ring exchange terms, where spin $k/2$ can mimic a $(k+1)-$body interaction (see Refs.~\onlinecite{schroeter1,greiterschroeter,anna} and references therein). In the context of possible applications in the context of quantum computation, this limitation is particularly severe, as only non-Abelian quasiparticles have the potential to act as topologically protected qubits, immune to any type of local perturbations such as disorder.

In this work we report on the discovery that non-Abelian Moore-Read (MR) and $Z_3$ Read-Rezayi (RR) phases naturally emerge as a consequence of realistic long-range interactions projected to a topological flat band, even at very high values of flux per plaquette where the lattice effects are significant. This result is in sharp contrast to what is found in earlier studies of Abelian FCIs, which are known to be weakened and ultimately destroyed by long-range interactions. A key difference between our work and previous studies is our choice of the Kapit-Mueller Hamiltonian \cite{kapitmueller} for the single-particle Chern band, which has a number of compelling features which make it particularly well suited for stabilizing exotic anyon states. While two very recent works have reported tentative evidence for the MR phase for realistic (short-range, two-body) interactions,\cite{cooperof,fcidmrg} our identification of this phase is arguably more compelling, using the full state-of-the-art toolbox of numerical techniques coupled with analytical considerations. Our observation of the $Z_3$ RR state for two-body interactions is entirely new and particularly important since the low-energy excitations include Fibonacci anyons which can be used to perform universal quantum computation (in contrast to the Majorana fermion excitations of the MR state, which cannot).

The remainder of this paper is organized as follows. In Sec.~\ref{model} we describe the Kapit-Mueller parent Hamiltonian for the lattice lowest Landau level (LLL) and describe various measures, including topological ground-state degeneracy, spectral flow, wave-function overlap, and particle-cut entanglement spectrum, which we use to verify the topological character of the states that we find through exact diagonalization. In Sec.~\ref{dipolar} we report on the results of our calculations for bosons with dipolar interactions in the Kapit-Mueller Hamiltonian and demonstrate the existence of non-Abelian anyon states. In Secs.~\ref{pseu} and \ref{optimize} we construct an analogous formulation of Haldane's pseudopotentials in our lattice Hamiltonian, and demonstrate how the structure of these pseudopotentials can be used to optimize the interactions to further increase the stability of non-Abelian ground states. In Sec.~\ref{candidate} we discuss how the model studied in our work might be implemented in realistic systems such as cold atoms or qubit arrays, and in Sec.~\ref{con} we offer concluding remarks.

\section{Hamiltonian and Methods}\label{model}

\subsection{Kapit-Mueller Hamiltonian}

We study $N_b$ interacting bosons on 2D square lattices with unit lattice spacing, with single-particle physics governed by the Kapit-Mueller Hamiltonian,\cite{kapitmueller}
\begin{eqnarray}\label{HKM}
H_0 &=& -\sum_{j \neq k} \of{ J \of{z_{j},z_{k}} a_{j}^{\dagger} a_{k} + {\rm H. c.}},
\end{eqnarray}
where
$J \of{z_{j},z_k} = J_0 W(z) e^{ (\pi/2) \of{z_{j} z^{*} - z_{j}^{*} z }  \phi }$,
$z\equiv  z_{j} - z_{k} \equiv x + i y $, and
$W(z)=\of{-1}^{1+ x+y+x y} e^{ -\of{\pi/2} \of{1-\phi} \abs{z}^{2}}$ in the symmetric gauge.
The hopping matrix elements of this Hamiltonian are complex, and the complex phase $e^{ (\pi/2) \of{z_{j} z^{*} - z_{j}^{*} z }  \phi }$ is nothing more than the symmetric gauge Peierls phase from a uniform magnetic flux which penetrates the lattice and has a strength of $\phi$ flux quanta per plaquette. We can rewrite this Hamiltonian in the Landau gauge via the transformation $a_{j}^{\dagger} \to a_{j}^{\dagger} \exp \of{-i \pi \phi x_j y_j }$. The Hamiltonian can be defined for any $0 \leq \phi < 1$, and in this work we study the rational flux densities $\phi = 1/q$ with $q=2,3,4,8$. The hopping matrix elements of (\ref{HKM}) are infinite ranged, but as they decay as a Gaussian, only a few terms beyond nearest-neighbor (NN) hopping need to be kept, and the Hamiltonian can be readily generalized to finite lattices of $L_1\times L_2$ sites with periodic boundary conditions by summing over lattice translations as described in Ref.~\onlinecite{kapitmueller}. A unit cell in the Landau gauge contains $q$ sites in the $x-$direction, so the lattice has $(qN_1)\times N_2$ sites if there are $N_1\times N_2$ unit cells. Following the standard conventions of the literature, the band filling fraction is defined as $\nu \equiv N_b/(N_1 N_2)$.

\begin{figure}
\includegraphics[width=\linewidth]{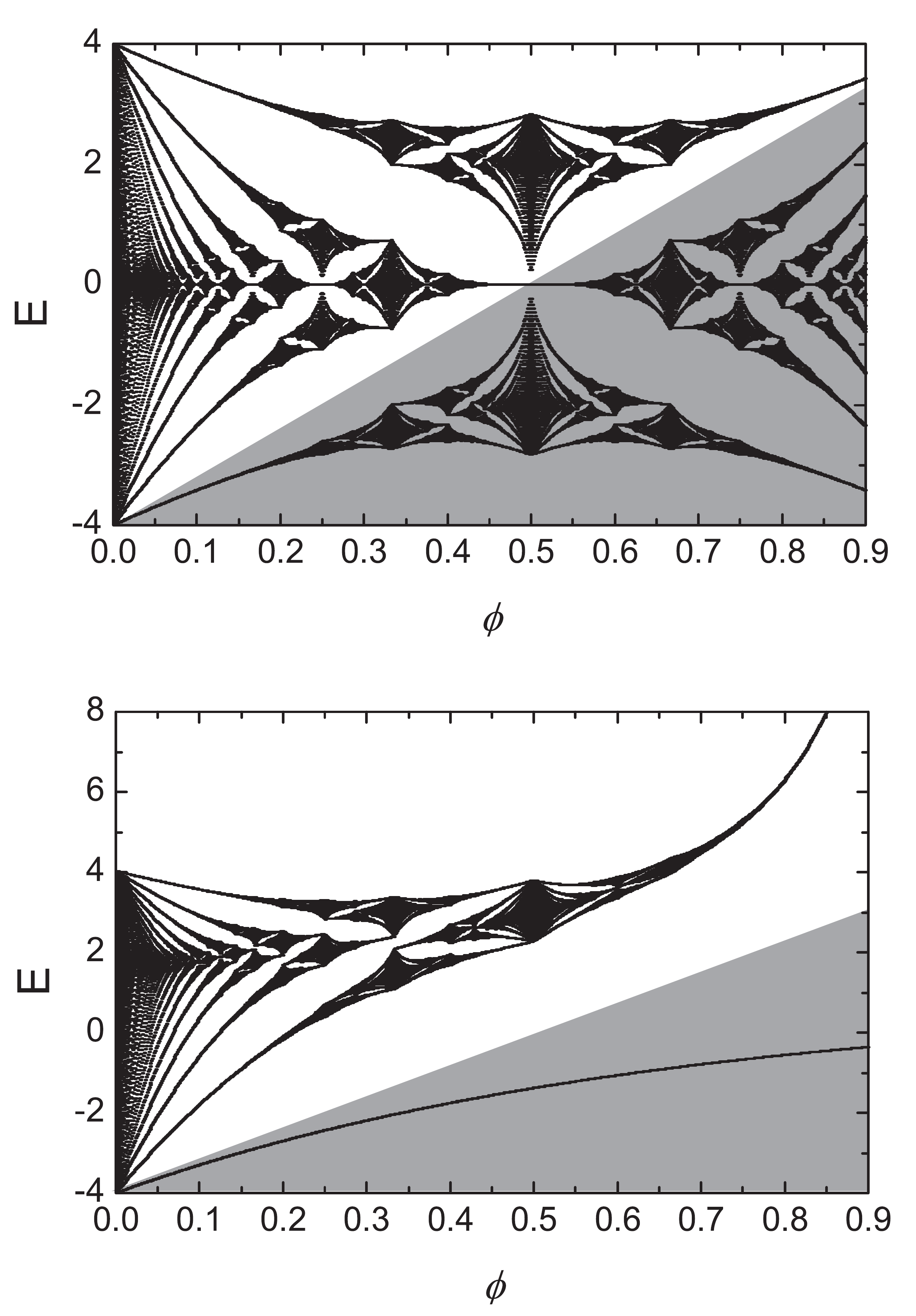}
\caption{(Color online) Flattening of the lowest band(s) with longer-ranged hopping on a square lattice. Top: Distribution of eigenvalues for the Hofstadter problem -- a nearest-neighbor hopping model on a square lattice in a uniform gauge field of strength $\phi$ quanta per plaquette, with the magnitude of the hopping matrix element set to unity. Bottom: The same distributions for the Kapit-Mueller Hamiltonian, (\ref{HKM}), with the nearest-neighbor hopping matrix element set to unity as well and a shift of the whole spectrum for a better comparison with the top figure. On an $L_1 \times L_2$ lattice with magnetoperiodic boundary conditions, the lowest $\phi L_1 L_2$ states (those within the shaded triangle) collapse to an exactly degenerate lowest Landau level. As $\phi \to 1$ the hopping becomes infinite ranged in this Hamiltonian, as seen in the diverging energy of the highest band. We only study $1/8 \leq \phi \leq 1/2$ in this work.}\label{NNvsLR}
\end{figure}

\begin{figure*}
\includegraphics[width=0.8\linewidth]{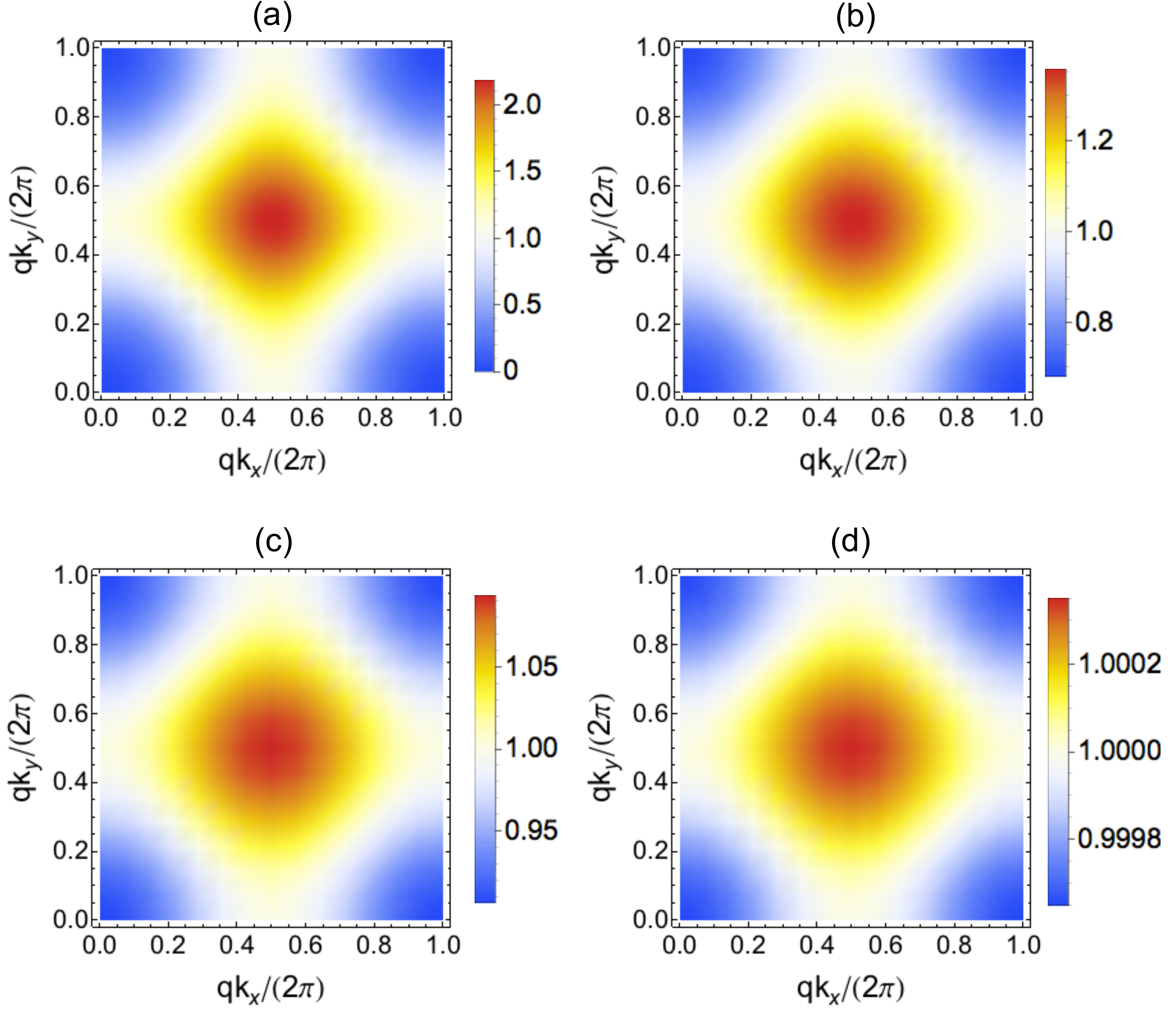}
\caption{(Color online) Berry curvature, rescaled by a multiplicative factor of $2\pi/q$, of the lowest band of the Kapit-Mueller Hamiltonian with (a) $\phi=1/2$, (b) $\phi=1/3$, (c) $\phi=1/4$, and (d) $\phi=1/8$, in $1/q$ of the Brillouin zone. The Berry curvature distribution is always inhomogeneous for finite $\phi$, but the variations quickly decay as $\phi\rightarrow 0$. See color bars for the rapidly changing scale.}
\label{Berry}
\end{figure*}

We study this Hamiltonian because its spectrum has an extensively degenerate ground state manifold which is exactly flat, separated from any excited bands by a large gap and spanned by lattice discretizations of the exact LLL wave functions of the continuum,\cite{kapitmueller} which are described by elliptic $\theta$ functions in the torus geometry that we study. These functions are evaluated over complex integer $z_j$ and the fraction of states in the lattice LLL is simply $\phi$. This analytical understanding of the wavefunctions allows us to exactly generalize a number of results found in the continuum LLL problem. In particular, for bosons at filling fraction $\nu = k/2$ with a $(k+1)-$body repulsive on-site interaction, the system's ground-state wave function is exactly given by the RR state of level $k$,\cite{readrezayi2} with a gap to both particle and hole excitations if the chemical potential is chosen to lie inside the many-body gap. We exploit this exactness to help confirm the topological character of the ground states with ranged two-body interactions, the focus of this work.

The dramatic flattening of the lowest band in this Hamiltonian can be seen in Fig.~\ref{NNvsLR}. It is important to note that, while the LLL wavefunctions have the same functional form as in the continuum and the magnetic length $l_{B} = 1/\sqrt{2 \pi \phi}$ can be less than a lattice spacing, the discretization of the LLL onto the lattice has important physical consequences. The lattice spacing itself introduces a second length scale into the system, and as a result the properties of a given state depend upon both the filling fraction $\nu$ and the flux density $\phi$. Unlike in the continuum LLL, two Kapit-Mueller systems with the same interactions but different flux densities can have topologically distinct ground states for the same $\nu$; for example, we were able to find $k=3$ RR ground states in bosons at $\nu = 3/2$ for $\phi= 1/4$ and $\phi = 1/3$, but not for $\phi = 1/2$. This occurs in part because the ground-state manifold of the Kapit-Mueller Hamiltonian is energetically flat but not ``Berry flat''; the gauge-invariant Berry curvature of the lowest band is positive definite across the magnetic Brillouin zone, but unlike the continuum, it is only uniform in the limit $\phi \to 0$ (Fig.~\ref{Berry}). Similarly, while the powers of $z_{j}^{n}$ or elliptic $\theta$ functions are exact ground states of Eq.~(\ref{HKM}), the discretization to the lattice breaks the rotational and continuous magnetic translational symmetries of the continuum and they are not naturally orthogonal. Consequently, Haldane pseudopotentials,\cite{haldaneexact,haldaneexacttorus} which are one of the most valuable tools in studying the continuum problem, cannot be generalized exactly to the lattice case. They can still be a valuable approximate tool in our model, however; we discuss an effective formulation of the Haldane pseudopotentials later in this work. In fact, the analogy with pseudopotentials leads us to introduce the new concept of "interaction flatness," which serves as an efficient guideline in the search for realistic models with stable non-Abelian FCIs.

\subsection{Topological degeneracies, trial wave functions, and entanglement spectrum}

The primary purpose of this work is to demonstrate the existence of non-Abelian anyon ground states of interacting bosons with ranged two-body interactions and with the single-particle physics described by Hamiltonian (\ref{HKM}). We do so by using numerical exact diagonalization to study small systems of up to $18$ interacting particles with magnetoperiodic boundary conditions, and to make the calculation numerically tractable we project onto the exactly flat lowest band of $H_0$ and diagonalize the interaction terms within the reduced basis. This approximation is good if the interaction potentials are weak compared to the band gap, which is $\sim 3-4 J_{NN}$ for the flux densities considered. We have three main tools at our disposal to demonstrate the anyon content of the resulting numerically generated ground states: topological degeneracies, overlap with the RR trial wave functions, and the particle-cut entanglement spectrum.

The anyon statistics of generic RR states require a ground-state degeneracy when the system is placed on a torus.\cite{wendagotto,greiterwen,oshikawa} For the $(k+1)-$body on-site interaction of lattice bosons this degeneracy is exact for any system size commensurate with the filling fraction, but for more general ranged interactions it is broken by finite-size effects, though the magnitude of this breaking is small compared to the excitation gap and is expected to vanish in the infinite-system limit. The $k=1,2,3$ states of bosons are two-, three-, and fourfold degenerate on the torus, and if the ground states of ranged interactions are in the identical topological phase, the same (approximate) ground-state degeneracy as in the RR states should be observed. While it is possible in principle that degeneracies in a ranged case could stem from separate effects which have nothing to do with the anyon content (such as charge density waves), we see no evidence of this occurring in our calculations.

The topological degeneracy of ground states should also be robust to twisted boundary conditions. For a many-body state $\Psi$, the twisted boundary condition in the $x(y)$ direction is defined as $\Psi(\mathbf{r}_j+L_{1(2)}\mathbf{e}_{x(y)})=\exp(i\Phi)\Psi(\mathbf{r}_j)$, where $\Phi$ is the magnetic flux inserted through the handle of the torus and $\mathbf{e}_{x(y)}$ is the unit vector in the $x(y)$ direction. $\Phi=0$ corresponds to the usual periodic boundary condition. In the spectral flow, i.e., the energy spectra as a function of $\Phi$, the ground states should never mix with excited levels.

In the continuum, the RR states of level $k$ \cite{readrezayi2} are the exact and unique (up to topological degeneracies) ground states of a $(k+1)-$body repulsive $\delta$ interaction in the LLL. As our lattice model, (\ref{HKM}), has the same single-particle wave functions as the continuum, the lattice discretizations of the RR states are exact ground states of the $(k+1)$-body on-site repulsion term $H_{\textrm{int}}=\frac{1}{(k+1)!}\sum_{i} \prod_{m=1}^{k+1} \of{n_{i} + 1 - m}$, where $n_i$ is the particle number at site $i$. It can be shown through conformal field theory,\cite{nayaksimon} Berry matrix analysis,\cite{nayakwilczek,georgiev,barabanzikos,prodan,kapitbraiding} or a plasma analogy \cite{laughlinoriginal,bondersongurarie} that the braid statistics of the anyon excitations of these states are Abelian for $k=1$, equivalent to Ising anyons or Majorana modes for $k=2$,\cite{mooreread} and $Z_{3}$ parafermions or Fibonacci anyons for $k=3$.

As we can numerically obtain the exact RR states in the lattice by diagonalizing the $(k+1)-$body on-site interaction, a simple tool at our disposal for predicting the topological character of the ground states for generic ranged interactions is the wave-function overlap. We first diagonalize the $(k+1)-$body on-site interaction to get a degenerate set of RR wave functions $\ket{\Psi^{i}_{\textrm{RR}}}$, and then diagonalize the ranged interactions to get a quasidegenerate set of ground states $\ket{\Psi^i}$. We then simply compute the total squared overlap $\frac{1}{k+1}\sum_{i,j=1}^{k+1}\abs{\left< \Psi^{i}_{\textrm{RR}} | \Psi^j \right> }^2$; if this quantity is close to unity, then the states are essentially identical and should have topologically identical excitations.

Entanglement measures can usually provide more insights into the topological order of the ground states than the overlap, which is only a single number that will vanish in the thermodynamic limit. Here we consider the entanglement spectrum that corresponds to the particle cut.\cite{lihaldane,sterregber} After dividing the whole system into two sets of particles, $A$ and $B$, which contains $N_A$ and $N_B=N_b-N_A$ bosons, respectively, we can obtain the reduced density operator of part $A$ of the ground state manifold by $\rho_A=\frac{1}{k+1}\textrm{Tr}_B(\sum_{i=1}^{k+1}|\Psi^i\rangle\langle\Psi^i|)$. The entanglement spectrum level is then defined as $\xi_i=-\ln \lambda_i$, where $\lambda_i$ is the eigenvalue of $\rho_A$. Due to translational invariance, each $\xi_i$ can also be labeled by the total momentum of part $A$. Considering that some particles are traced out artificially, the particle-cut entanglement spectrum should contain information about the quasihole excitations, which is a fingerprint of topological order in the ground-state manifold.

\begin{figure*}
\includegraphics[width=\linewidth]{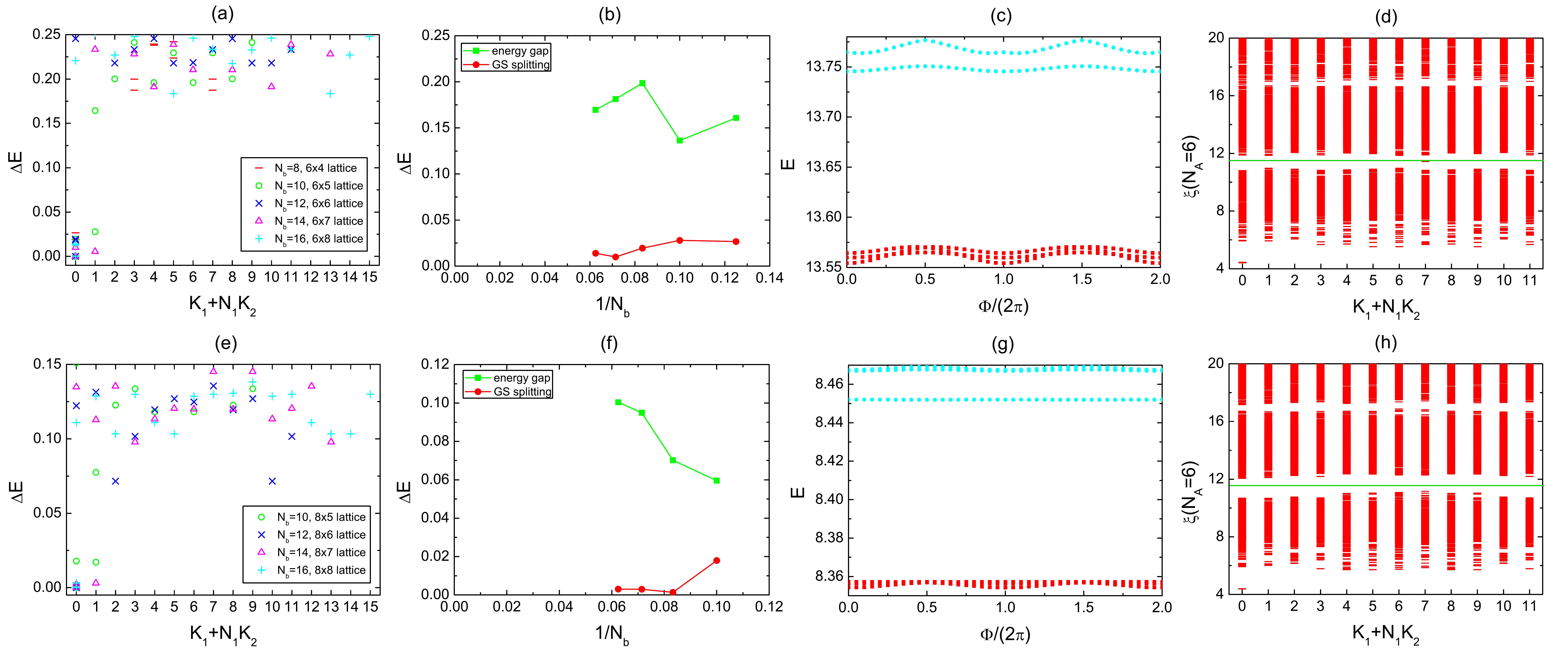}
\caption{(Color online) Numerical evidence for bosonic Moore-Read states at $\nu=1$ for (a-d) $\phi=1/3$ and (e-h) $\phi=1/4$, with two-body on-site and dipolar interactions, (\ref{int}). We choose $U=1.0$ for $\phi=1/3$ and $U=-0.6$ for $\phi=1/4$. (a, e) Energy spectra for $N_b$ bosons in a square lattice of $qN_1\times N_2$ sites. (b ,f) Finite-size scaling of both energy gap and ground-state splitting. The scaling behavior depends not only on the system size but also on the aspect ratio $L_1/L_2$ of the lattice. (f) The gap even goes up when the system size increases because the sample becomes more isotropic. (c, g) The $x-$direction spectral flow for $N_b=14$ bosons, $N_1=2$, and $N_2=7$. (d, h) Particle-cut entanglement spectra for $N_b=12$ bosons, $N_1=2$, and $N_2=6$. The number of levels below the gap (solid line) is $3430$, matching the quasihole excitation counting of MR states.}
\label{bosonMR}
\end{figure*}

\section{Non-Abelian States of Dipolar Bosons}\label{dipolar}

We begin the discussion of our results by considering the experimentally realistic case of bosons with two-body on-site and dipolar interactions,
\begin{eqnarray}\label{int}
H_{\textrm{int}}= \frac{U}{2} \sum_i n_i(n_i-1)+\frac{V}{2}\sum_{i\neq j}\frac{n_i n_j}{|\mathbf{r}_i-\mathbf{r}_j|^3},
\end{eqnarray}
where $i$ and $j$ are site indices and $V$ is set equal to $1$ in this section. We implicitly assume that $\abs{V} \ll J_0$ so that we need consider only the lowest band of $H_0$ in our calculations. We truncate the dipolar interaction by only considering the nearest distance between two sites on the torus when evaluating the dipolar matrix elements (i.e., we do not sum over periodic contributions). We then scan the values of $U$ for fixed $\phi$ to search for the features described in the previous section, which tell us when the ground states of (\ref{int}) are in the same topological phase as the RR states.

We first focus on the filling fraction $\nu=1$ ($k=2$), where the RR state is also called the MR state (for the remainder of this work, RR state will refer only to the $k=3$ case). Compelling evidence, as displayed in Fig.~\ref{bosonMR} and Table~\ref{overlapMR}, demonstrates that the ground states of (\ref{int}) are indeed in the MR phase for flux densities as high as $\phi=1/4$ and $\phi=1/3$. After choosing appropriate $U$, there are always three quasidegenerate ground states for each system size that we study, and they are separated from the excited levels by an energy gap (defined as the difference between the lowest excited level and the highest ground level) which is significantly larger than the ground-state splitting [Figs.~\ref{bosonMR}(a) and \ref{bosonMR}(e)]. A finite-size scaling analysis strongly suggests that the gap will survive in the thermodynamic limit [Figs.~\ref{bosonMR}(b) and \ref{bosonMR}(f)]. By calculating the spectral flow, we find that the topological degeneracy is robust to the twisted boundary conditions, i.e., the ground manifold never mixes with the excited levels [Figs.~\ref{bosonMR}(c) and \ref{bosonMR}(g)]. The total squared overlaps between the ground states of (\ref{int}) and the exact MR states are highly nontrivial (Table~\ref{overlapMR}). The particle-cut entanglement spectra show clear entanglement gaps, below which the number of levels is the same as the quasihole excitation counting of MR states [Figs.~\ref{bosonMR}(d) and \ref{bosonMR}(h)]. These results conclusively confirm the existence of the MR phase in the presence of on-site and ranged dipolar interactions, (\ref{int}), at high flux densities.

\begin{figure*}
\includegraphics[width=0.7\linewidth]{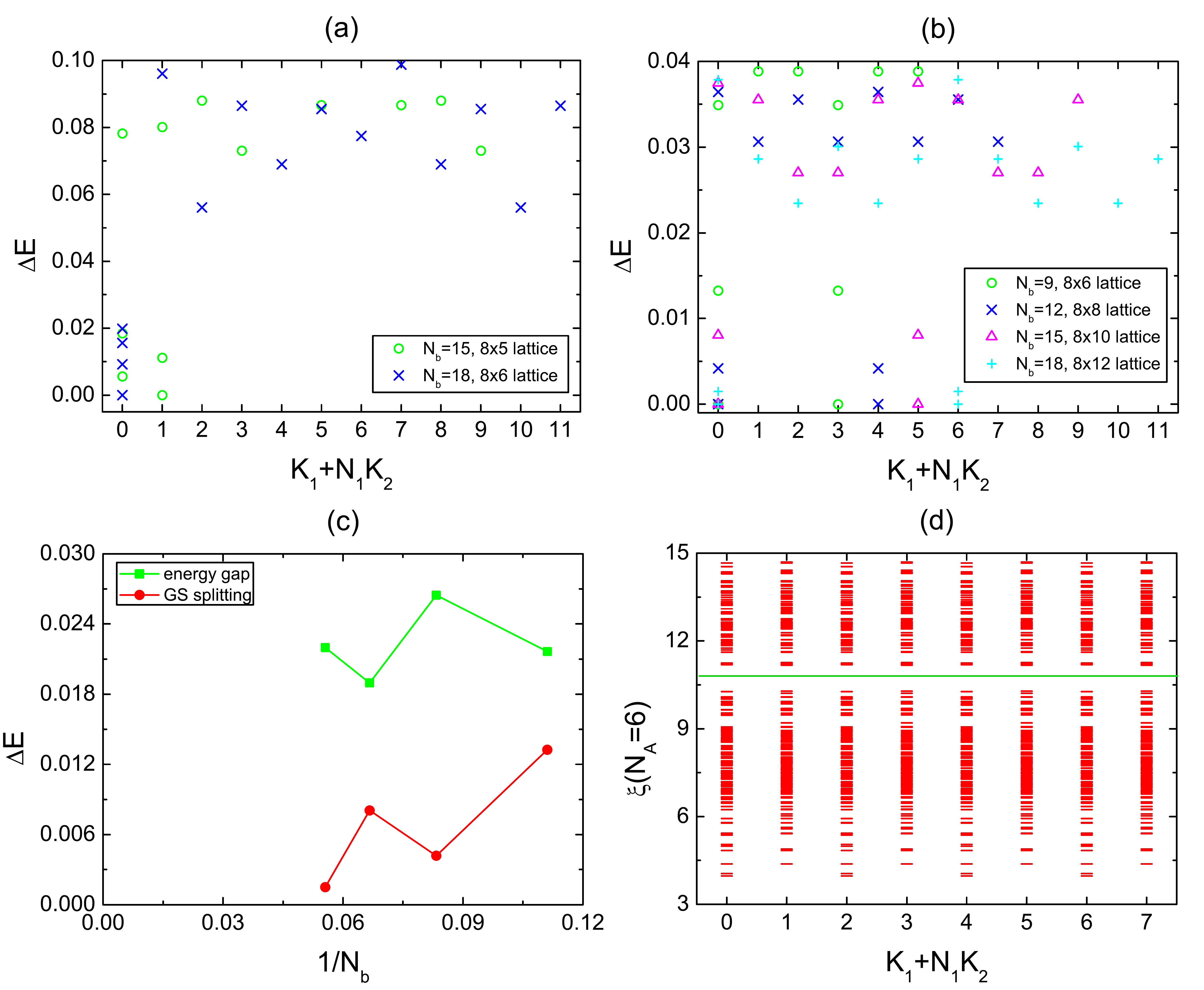}
\caption{(Color online) Numerical evidence for bosonic Read-Rezayi states at $\nu=3/2$ for (a) $\phi=1/4$ and (b-d) $\phi=1/8$, with two-body on-site and dipolar interactions, (\ref{int}). We choose $U=-0.6$ for $\phi=1/4$ and $U=-2.7$ for $\phi=1/8$. (a, b) Energy spectra for $N_b$ bosons in the square lattice of $qN_1\times N_2$ sites. (c) Finite-size scaling of both energy gap and ground-state splitting for $\phi=1/8$. (d) Particle-cut entanglement spectra for $N_b=12$ bosons, $N_1=1$, $N_2=8$, and $\phi=1/8$. The number of levels
below the gap (solid line) is $884$, matching the quasihole excitation counting of RR states. }
\label{bosonRR}
\end{figure*}

Compared with the MR phase, the RR phase at $\nu=3/2$ ($k=3$) is more fragile and sensitive to details such as the precise form of the interactions and the lattice size. The need to go to a slightly lower flux as well as the higher filling fraction complicates the numerical simulations in the sense that the number of available lattice samples is decreased by the increasing size of the computational basis, making faithful finite-size extrapolations harder to achieve. Nevertheless, we still observe very encouraging evidence that the RR phase might be stabilized at a flux density as high as $\phi=1/4$ using the dipolar interaction with a suitably tuned on-site interaction. In Fig.~\ref{bosonRR}(a), we show the energetic data for $\phi=1/4$ with the interaction, (\ref{int}), in which we use the same $U$ as for the MR case. The topological degeneracy supports that the ground states are in the RR phase. However, the disappearance of a clear fourfold quasidegeneracy for $N_b<15$, which may be caused by the less isotropic aspect ratio $L_1/L_2$, makes it difficult to do a reliable finite-size scaling of the energy gap. The total squared overlap between the ground states and the exact RR states is lower than that in the MR case, due to which the entanglement gap in the particle-cut entanglement spectra is invisible for large $N_A$. Nevertheless, the overlap reaches $0.8220$ for $15$ bosons, which is very encouraging (Table~\ref{overlapRR}). On general grounds, we expect that the evidence of RR states will become more compelling for smaller $\phi$. The $\phi=1/8$ results displayed in Figs.~\ref{bosonRR}(b) and \ref{bosonRR}(d) and Table~\ref{overlapRR} confirm this expectation and conclusively establish the existence of the RR phase for realistic interactions in the Kapit-Mueller lattice model at a relatively high flux density.

\begin{table}
\caption{Total squared overlaps between the ground states of two-body interactions (\ref{int}) at $\nu=1$ and the exact Moore-Read states for $\phi=1/3$ and $\phi=1/4$. Lattice sizes and interaction parameters are the same as in Fig.~\ref{bosonMR}. N/D: There is no clear threefold quasidegeneracy for this sample.\label{overlapMR}}
\begin{ruledtabular}
\begin{tabular}{ccccc}
&$N_b=8$&$N_b=10$&$N_b=12$&$N_b=14$\\
\hline
$\phi=1/3$&$0.9142$&$0.9349$&$0.9012$&$0.8806$\\
$\phi=1/4$&N/D&$0.9279$&$0.9280$&$0.8949$
\end{tabular}
\end{ruledtabular}
\end{table}

\begin{table}
\caption{Total squared overlaps between the ground states of two-body interactions (\ref{int}) at $\nu=3/2$ and the exact Read-Rezayi states for $\phi=1/4$ and $\phi=1/8$. Lattice sizes and interaction parameters are the same as in Fig.~\ref{bosonRR}. N/D: There is no clear fourfold quasidegeneracy for this sample. \label{overlapRR}}
\begin{ruledtabular}
\begin{tabular}{cccc}
&$N_b=9$&$N_b=12$&$N_b=15$\\
\hline
$\phi=1/4$&N/D&N/D&$0.8220$\\
$\phi=1/8$&$0.8682$&$0.9025$&$0.8843$
\end{tabular}
\end{ruledtabular}
\end{table}

\section{Pseudopotential analogy}\label{pseu}

\begin{figure*}
\includegraphics[width=0.7\linewidth]{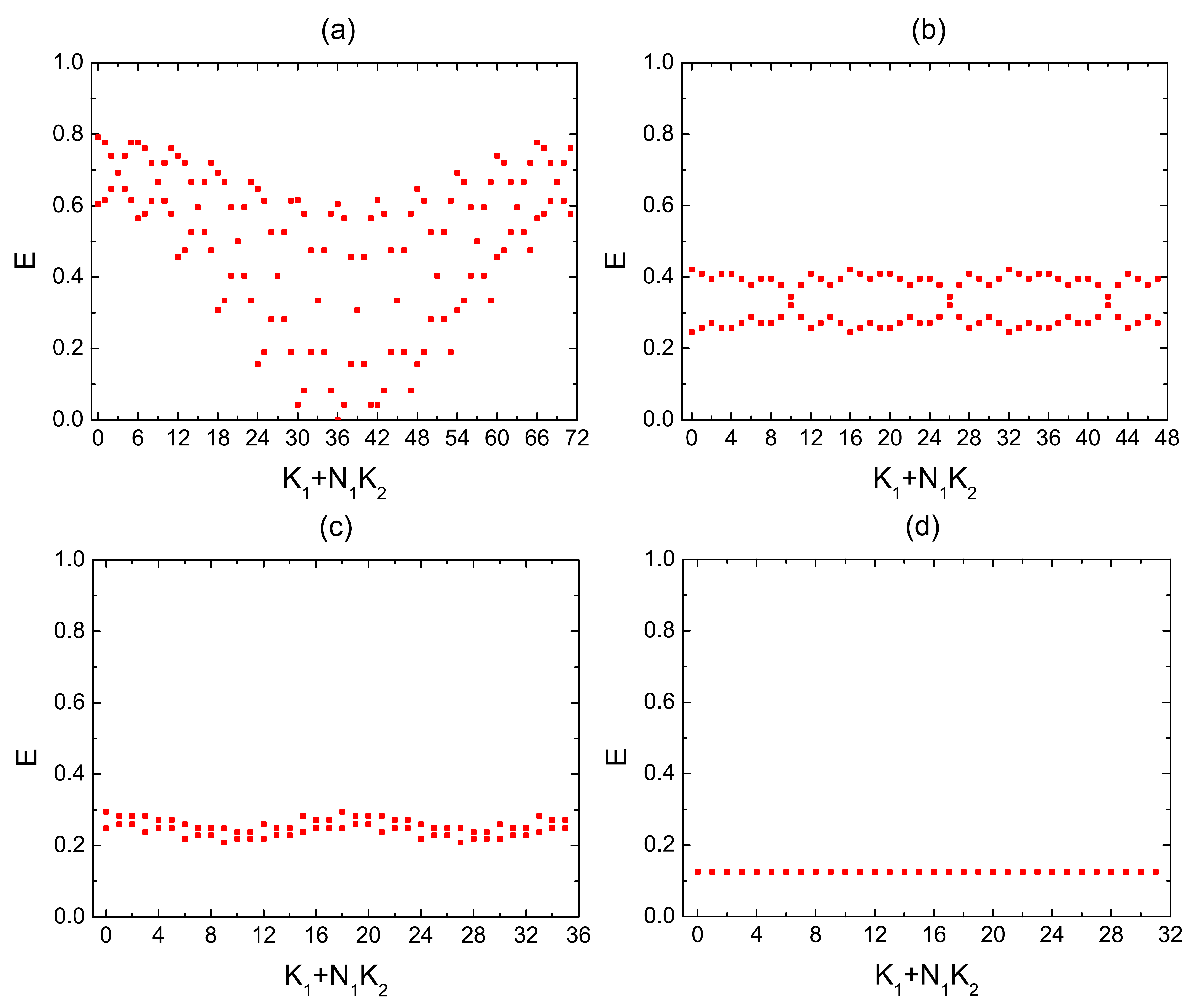}
\caption{(Color online) Two-particle spectrum for the on-site interaction on a $12\times12$ lattice for (a) $\phi=1/2$, (b) $\phi=1/3$, and (c) $\phi=1/4$ and on a $16\times16$ lattice for (d) $\phi=1/8$. }
\label{onsitetwobody}
\end{figure*}

In this section we corroborate the numerical evidence for non-Abelian FCIs in the lattice by drawing connections to
Haldane's pseudopotentials;\cite{haldaneexact} in particular, we discuss to what extent the much simpler problem of two-particles interacting within a Chern band \cite{lauchli} can guide the search for exotic FCI phases.

In continuum Landau levels, Haldane's pseudopotentials offer a very useful framework for understanding the phase diagram of interacting particles by providing an expansion of any interaction in terms of parent Hamiltonians for various model states. As we shall see, the insights that we gather also provide efficient guiding principles for engineering optimal models stabilizing non-Abelian FCIs, and it is precisely these principles which led us to the results in the previous section.

In the original problem of the continuum LLL, the pseudopotentials were formulated as
 \begin{eqnarray}H=\sum_{i<j}\sum_{m=0}^{\infty}\mathcal V_mP_m^{ij},\label{pseudohamiltonian}\end{eqnarray}
where $P_m^{ij}$ projects onto a state where particles $i,j$ have
relative angular momentum $m$ and $\mathcal V_m$ are the pseudopotential
parameters, which are real numbers determined by the functional form of the
interaction (and by the form factor of the pertinent Landau level). Any two-body potential interaction $V(\mathbf{r})$ can be written as a sum of pseudopotential terms, with positive $\mathcal{V}_m$ corresponding to repulsive channels and negative $\mathcal{V}_m$ corresponding to attractive ones. More generally, the pseudopotentials project onto components of the many-body wave function with certain vanishing properties and, as such, can be defined in generic geometries such as the torus.\cite{haldaneexacttorus} Up to a $q-$fold topological degeneracy, the Laughlin states at filling fraction $\nu=1/q$ are the unique zero-energy ground states with $\mathcal V_m>0, m\leq q-2$ and  $\mathcal V_m=0, m> q-2$. For bosons (fermions) only the even (odd) pseudopotentials occur in the Hamiltonian due to the (anti-)symmetry of the many-body wave functions.

When the pseudopotential distribution is sufficiently peaked at small $m$, as in the case in the lowest electronic Landau levels of conventional semiconductor heterostructures, it is well known that a hierarchy of Abelian fractional quantum Hall states is realized.\cite{haldaneexact, Halperinhierarchy,Jain89} The basic physics of this hierarchy is that, in the vicinity of a Laughlin fraction, $\nu=1/q$, the fractionally charged quasiparticles themselves experience an effective repulsive short-range interaction and thereby form Laughlin-like liquids on their own. When the pseudopotential distribution is less peaked, as in higher Landau levels, pairing or even clustering may instead be favored. In particular, this is believed to happen in the half-filled second Landau level ($\nu=5/2$), where the quasiparticles undergo $p$-wave pairing akin to superconductivity (of spinless particles).\cite{mooreread,readgreen} Even more exotically, at Landau level filling $\nu=12/5$ (and $\nu=13/5$), there is some evidence \cite{RRnumerics} that the nonpeaked pseudopotential distribution may lead to three-particle clustering and the resulting formation of RR states.

Building on the ideas of L\"auchli {\it et al.},\cite{lauchli} who realized that there is a natural generalization of the pseudopotential arguments to the lattice via the two-particle problem projected to a Chern band, we now rationalize our findings on non-Abelian FCIs. In fact, the pseudopotential analogy turns out to be particularly enlightening in the context of the Kapit-Mueller model, as we elaborate on in the following.

Let us first consider the case of on-site interactions. The two-particle spectrum, $\{E_{n}(\mathbf K)|E_{n}(\mathbf K)\geq E_{n+1}(\mathbf K), n=0,1,2,...\}$, of this problem is displayed in Fig.~\ref{onsitetwobody} for various flux densities. Notably there are two nonzero eigenvalues, $E_{0}(\mathbf K)$ and $E_{1}(\mathbf K)$, in each momentum sector.\footnote{An exception is the $\phi=1/2$ case, where one of the sectors has only a single nonzero eigenvalue. This is, however, a nongeneric feature---for instance, the second level occurs for twisted boundary conditions---and it does not influence the subsequent discussion.} These eigenvalues are split and also depend on the center-of-mass momentum, $\mathbf K$. For smaller $\phi$ these ``imperfections'' decrease, and in the limit $\phi\rightarrow 0$ they vanish. In this limit, the symmetry of the problem is enhanced and we can identify the spectrum with that of two particles projected to a Landau level, where an identical spectrum is obtained for the $delta-$function potential, which is precisely the zeroth pseudopotential, with strength $\mathcal{V}_0$. Hence we can identify $\mathcal{V}_0 =E_{0}(\mathbf K)$ [=$E_{1}(\mathbf K)$] in our lattice model as $\phi\rightarrow 0$.

This observation suggests that the replacement
\begin{eqnarray}
\mathcal V_0\rightarrow \mathcal V_0(\mathbf K)=[E_{0}(\mathbf K)+E_{1}(\mathbf K)]/2,
\label{v0k}
\end{eqnarray}
where the parameters $\mathcal V_0(\mathbf K)$ can be directly obtained by diagonalizing the problem of two particles projected to the pertinent (lowest) Chern band, is suitable for describing the full many-body problem in the lattice at general $\phi$. The validity of this replacement is corroborated by the fact that the Laughlin state at $\nu=1/2$ is indeed the exact gapped ground state for on-site repulsion in the Kapit-Mueller model at any $\phi$.

\begin{figure*}
\includegraphics[width=0.7\linewidth]{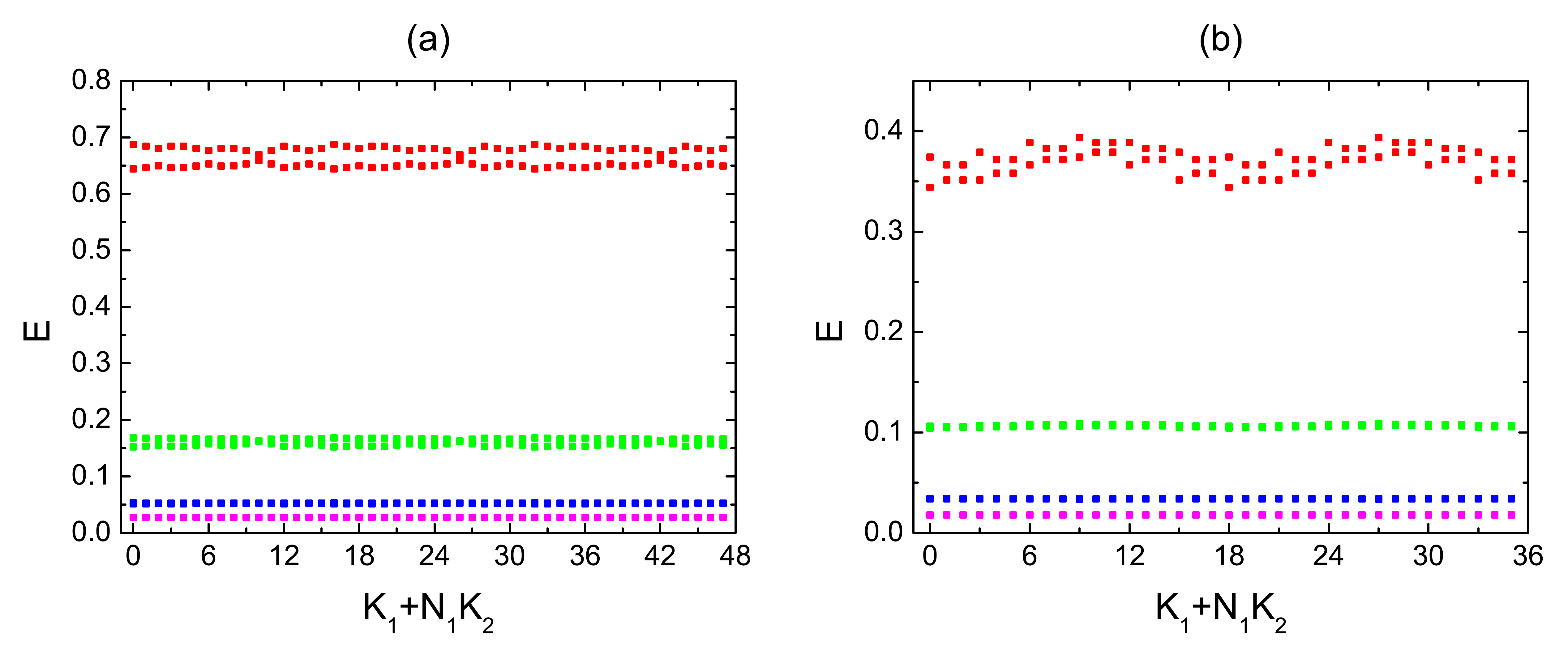}
\caption{(Color online) Two-particle spectrum on a $12\times12$ lattice for (a) $\phi=1/3, N_1\times N_2=4\times12$ and (b) $\phi=1/4, N_1\times N_2=3\times12$, with dipolar interactions (\ref{int}). We choose the on-site interaction as $U=1.0$ for $\phi=1/3$ and $U=-0.6$ for $\phi=1/4$. Only the largest eight levels are shown. }
\label{twobody}
\end{figure*}

More boldly, we suggest that the generalization
\begin{eqnarray}
\mathcal V_m\rightarrow \mathcal V_m(\mathbf K)=[E_{2m}(\mathbf K)+E_{2m+1}(\mathbf K)]/2,
\label{vmk}
\end{eqnarray}
is appropriate for arbitrary $m$ (cf. Ref.~\onlinecite{lauchli} for a complementary, heuristically motivated, argumentation). Clearly, it is easy to come up with counterexamples, such as ``hollow-core'' interactions where the identification, (\ref{vmk}), needs modification. Nevertheless, this heuristic serves as an efficient guideline for the class of interactions we consider in this work, and as a {\it first guiding principle}, we look for interactions which lead to average generalized pseudoptentials,
\begin{eqnarray}
\tilde{ \mathcal V}_m=\frac{1}{N_1N_2}\sum_{\mathbf K}\mathcal V_m(\mathbf K) ,
\label{average}
\end{eqnarray}
similar to the $\mathcal{V}_{m}$'s where non-Abelian phases are know to occur in the continuum (whenever such information is available). In Table \ref{dipolar} we compare the leading $\tilde{ \mathcal V}_m$'s, extracted from the data in Fig.~\ref{twobody}, for optimal choices of the on-site interaction $U$, for the dipolar interactions (\ref{int}), with optimal pseudopotential parameters, $\mathcal V_m$, in the continuum. The similarities are rather striking even though the precise values depend slightly on the aspect ratio $L_1/L_2$ of the sample and the definition of ``optimal,'' i.e., if the parameters are chosen so as to optimize the overlaps or whether an optimized ground-state degeneracy is the goal. Our goal is to observe excellent degeneracy for each system size, rather than to pursue the largest overlap for one fixed system size.

\begin{table}
\caption{Pseudopotential parameters, (\ref{average}), extracted from Fig.~\ref{twobody}, compared with those for which the ground states of two-body interactions, (\ref{int}), have the largest overlap with the exact MR states in the continuum. Continuum results were obtained for $10$ bosons on a sample with $L_1/L_2=1$.\cite{seki} Here we rescale the energy so that $\tilde{\mathcal V}_0=1$. \label{pp}}
\begin{ruledtabular}
\begin{tabular}{cccc}
&$\phi=1/3,U=1.0$&$\phi=1/4,U=-0.6$&continuum\\
\hline
$\tilde{ \mathcal V}_0$&$1$&$1$&$1$\\
$\tilde{\mathcal V}_2$&$0.2423$&$0.2869$&$0.3$\\
$\tilde{\mathcal V}_4$&$0.0787$&$0.0916$&$0.0938$\\
$\tilde{\mathcal V}_6$&$0.0414$&$0.0482$&$0.0492$
\end{tabular}\label{dipolarpp}
\end{ruledtabular}
\end{table}

We note that although $\mathcal V_m(\mathbf K)$ varies as a function of the center-of-mass momentum $\mathbf K$, this should not affect the projector properties of the corresponding Hamiltonian (as manifest in the $m=0$ case discussed above). However, for states that are not exact zero modes of any two-body Hamiltonian, the variation with $\mathbf K$ may have a crucial influence and provides a hint as to why non-Abelian FCIs have proven so hard to realize for two-body interactions. Consequently, as the {\it second, equally important, guiding principle} our search for non-Abelian states in bosons with long-ranged interactions is guided by the conjecture that non-Abelian states will be most stable when the pseudopotential distributions $\mathcal{V}_{m} (\mathbf{K})$ are nearly flat and independent of $\mathbf{K}$, as they are in the continuum LLL. While this idea is only heuristic, we can justify it in part by noting that the gapped FCI states effectively fill the LLL, and involve contributions from every momentum wavevector allowed by the boundary conditions. If the interaction pseudopotentials have a broad spread in $\mathbf{K}$, then it is easy to imagine low-energy modes arising from small variations in the relative occupations of momenta in the ground state, whereas if the pseudopotentials are flat, the variation in energy arising from variations in the ground-state momenta should be comparatively larger, increasing the ground-state gap. In other words, we expect that ``interaction flatness'' should make an important contribution to the stability of FCI states, just as Berry flatness and band flatness do in both Chern bands and the continuum LLL.

\begin{figure*}
\includegraphics[width=0.75\linewidth]{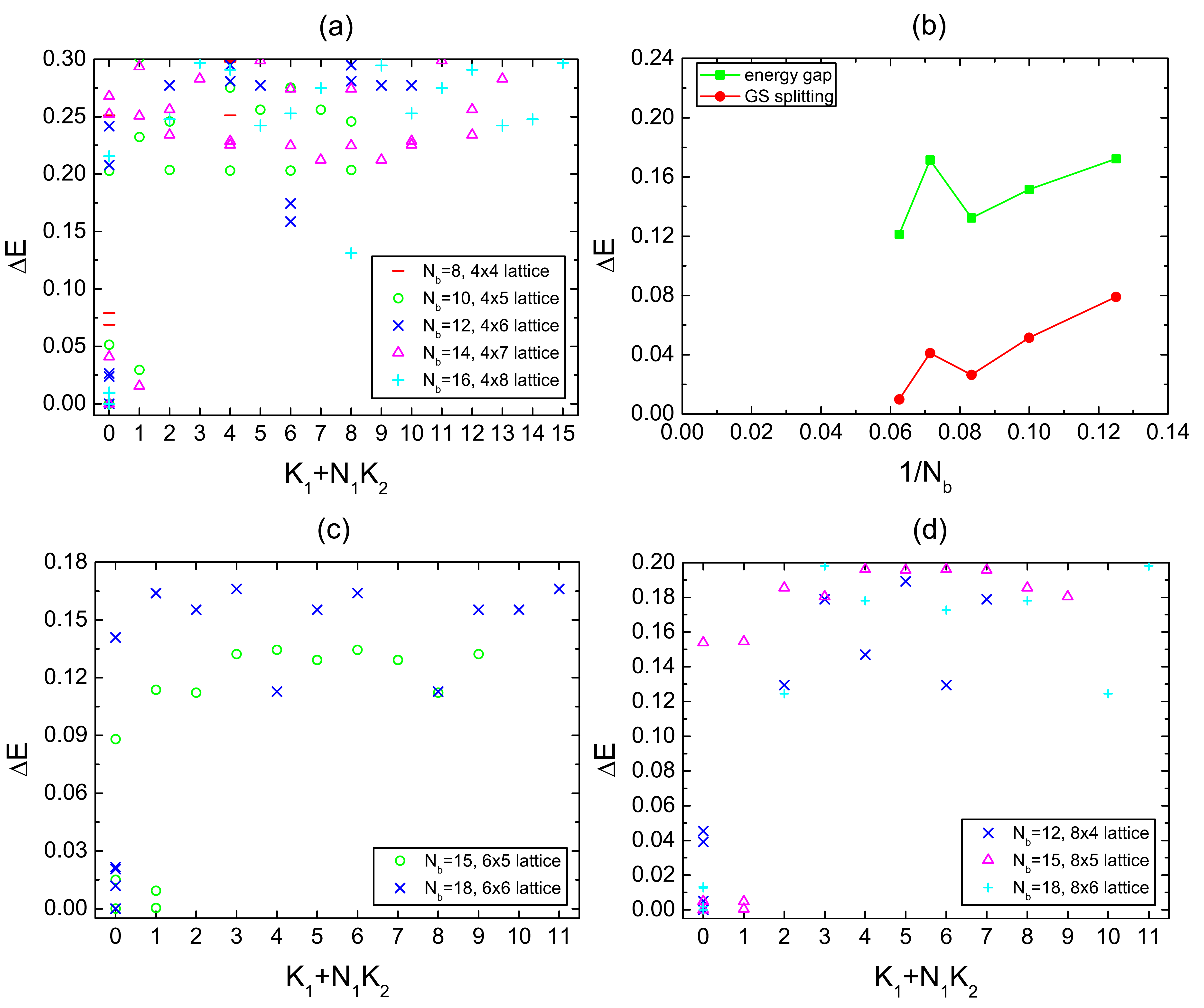}
\caption{(Color online) Numerical evidence for bosonic non-Abelian states at $\nu=1$ and $\nu=3/2$, with two-body interactions, (\ref{finet}).
(a) Energy spectra and (b) the finite-size scaling of the energy gap and ground-state splitting at $\nu=1$ for $\phi=1/2$. We choose $n_{{\rm max}}=2$ with $V_1=0.65$ and $V_2=0.4$.
(c) Energy spectra at $\nu=3/2$ for $\phi=1/3$. We choose $n_{{\rm max}}=4$ with $V_1=1$, $V_2=V_3=0.55$, and $V_4=0.35$.
(d) Energy spectra at $\nu=3/2$ for $\phi=1/4$. We choose $n_{{\rm max}}=4$ with $V_1=0.9$, $V_2=0.7$, $V_3=0.6$, and $V_4=0.2$.}
\label{bosonfinetune}
\end{figure*}

While the requirement of interaction flatness is purely a conjecture and the existence of exact, zero-energy ground states in the Kapit-Mueller Hamiltonian for $(k+1)-$ body on-site interactions, even at flux densities where the on-site interaction has a broad energy range when projected to the LLL, is an important counterexample, it, along with the first guiding principle, provides an extremely useful tool when searching for non-Abelian states. Namely, while the computational cost of exact diagonalization increases exponentially with system size, for any well-defined interaction $V(\mathbf{r})$ the computational cost of obtaining two-particle spectra such as those shown in Figs.~\ref{onsitetwobody} and \ref{twobody} is essentially negligible, so it is easy to conduct a brute force search in the two-particle spectra to find combinations of interactions which lead to nearly flat pseudopotentials. As mentioned above, knowledge of the optimal ratios of the $\mathcal{V}_{m}$ from continuum LLL studies \cite{Morf,cooperwilkin,RezayiHaldane2000,RRnumerics,rezayireadcooper} of the desired phase should be further employed to guide this search. Once an ``optimal'' set of interaction terms is determined, the full many-body Hamiltonian can be diagonalized to determine whether or not those interactions can indeed stabilize the target non-Abelian state, and small variations in the interaction terms can then be studied through exact diagonalization to search for local maxima in the energy gap and local minima in the finite-size degeneracy breaking. We employed precisely this procedure to find most of the non-Abelian states reported in this work, and we believe that it could be an important guiding principle for future searches for exotic states in other Chern band Hamiltonians. In the case of the dipolar interactions studied in Sec.~\ref{dipolar}, the search for optimal interactions to stabilize non-Abelian FCIs was conducted over the single parameter $V/U$, and as shown in Fig.~\ref{twobody} the two-particle spectra are indeed flatter than those in Fig.~\ref{onsitetwobody} for the optimal parameter. As we show in the next section, the study of the two-particle spectra also allows for much more general searches over many tuning parameters. Such searches would be impossible with brute force exact diagonalization for large systems, but are easier in the two-particle pseudopotential framework and can dramatically increase the stability of the higher order MR and RR states.

We remark that, as done earlier in the continuum,\cite{multipseudo} the pseudopotential concept can also be extended to multibody interactions. Analogously with the two-body case, this provides a rationale for the fact that the on-site $\of{k+1}-$body repulsive interaction acts as a parent Hamiltonian for the $Z_k$ RR states in the Kapit-Mueller model. Even though other lattice models with flat Chern bands do not have the benefit of exact model states for short-range interactions we expect that our insights carry over rather directly to more generic models, in agreement with the heuristic arguments in Ref.~\onlinecite{lauchli}.

For completeness we also wish to mention that there is another approach to extract effective pseudopotentials in Chern bands due to Lee \textit{et al.} \cite{leethomaleqi,leeqi}. We do not use their formalism in this work, in part because they do not consider the momentum dependence of the $\mathcal{V}_m$ directly. Yet another approach, whereby continuum pseuopotentials were constructed to fit various FCIs, was considered by Wu \textit{et~al.} \cite{wubloch}. Although conceptually pleasing and helpful in confirming the nature once an FCI phase is identified, the approach of Wu \textit{et al.} does not provide guidance in the search for lattice models harboring non-Abelian FCI phases.

\section{Optimizing Interactions to Stabilize Non-Abelian States}\label{optimize}

In Sec.~\ref{dipolar}, we show how to stabilize MR and RR states by experimentally realistic, infinite-ranged interactions. The pseudopotential analysis in Sec.~\ref{pseu} suggests that it is also possible to reach the MR and RR phases by fine-tuning finite-range two-body interactions as long as the two-particle spectra are relatively flat and the average generalized pseudopotentials $\tilde{ \mathcal V}_m$ are appropriate. In fact, by allowing ourselves to tune more interaction parameters we are able to stabilize non-Abelian FCIs at even higher flux densities. Specifically, we consider a combination of the on-site term and $n$-th NN terms [$n=1$ corresponds to the NN interaction, $n=2$ corresponds to the next-nearest-neighbour (NNN) interaction, etc.],
\begin{eqnarray}
H_{\textrm{int}}=\frac{U}{2} \sum_i n_i(n_i-1)+\sum_{m=1}^{n_{{\rm max}}} V_m\Big(\sum_{(i,j)\in \mathcal{N}_m} n_i n_j\Big),
\label{finet}
\end{eqnarray}
where $\mathcal{N}_m$ is the set of all $m$-th NN sites and $U$ is set equal to $2$ in this section. $n_{{\rm max}}$ indicates the range of the interactions, and we choose $n_{{\rm max}}$ as small as possible. By choosing appropriate $V_m$, we obtain stronger evidence of the non-Abelian phases, even for system sizes where these phases are absent with the combination of on-site and dipolar interactions, (\ref{int}).

Let us first consider the MR phase for $\phi=1/2$. As discussed earlier we were not able to identify the necessary threefold quasidegeneracy with the single-parameter family of interactions given in Eq. (\ref{int}) for this high flux density.
However, we find that an appropriate choice of more generic interactions, (\ref{finet}), can stabilize the MR phase [Figs.~\ref{bosonfinetune}(a) and \ref{bosonfinetune}(b)]. Already by keeping on-site and NN interactions, we are able to identify a region in parameter space where an excellent threefold quasidegeneracy for each studied system size, ranging from $8$ to $16$ bosons, occurs. Importantly, a finite-size scaling analysis suggests that the energy gap can survive in the thermodynamic limit in this regime [Fig.~\ref{bosonfinetune}(b)]. The overlap results listed in Table~\ref{overlapMR2} also clearly corroborate our belief that the ground states are indeed representing the MR phase.

By tuning more interaction parameters, we have also obtained evidence of the RR phase for $\phi=1/3$ and increased its stability at $\phi=1/4$. For $\phi=1/3$, the RR phase is not observed with the combination of on-site and dipolar interactions, (\ref{int}). However, fine-tuning the interactions of the form (\ref{finet}) with $n_{{\rm max}}=4$ (a four-dimensional search made possible by the flat pseudopotential heuristic of the previous section), we observe fourfold degeneracy for $15$ and $18$ bosons [Fig.~\ref{bosonfinetune}(c)]. For $\phi=1/4$, we find stronger evidence of the RR phase compared with that in Sec.~\ref{dipolar}. The fourfold degeneracy appears for more samples, and the energy gap for the two largest system sizes ($N_b=15$ and $N_b=18$) is roughly twice as large as that in Fig.~\ref{bosonRR}(a), while the ground-state splitting is smaller [Fig.~\ref{bosonfinetune}(d)]. The total squared overlaps in Table~\ref{overlapRR2} also show a large improvement compared with those in Table~\ref{overlapRR}. As an example, the squared overlap for $N_b=15$ bosons increases from $0.8220$ to $0.8849$.

In each of the above cases we find it particularly encouraging that the topological degeneracies generically become more pronounced at larger system sizes. In fact, we did not observe the opposite behavior in any of the cases discussed in this work; i.e., we did not observe the approximate degeneracies to weaken or disappear in the large samples.

\begin{table}
\caption{Total squared overlaps between the ground states of two-body interactions (\ref{finet}) at $\nu=1$ and the exact Moore-Read states for $\phi=1/2$. Lattice sizes and interaction parameters are the same as in Fig.~\ref{bosonfinetune}(a).
\label{overlapMR2}}
\begin{ruledtabular}
\begin{tabular}{ccccc}
&$N_b=8$&$N_b=10$&$N_b=12$&$N_b=14$\\
\hline
$\phi=1/2$&$0.8891$&$0.8509$&$0.8101$&$0.8269$
\end{tabular}
\end{ruledtabular}
\end{table}

\begin{table}
\caption{Total squared overlaps between the ground states of two-body interactions (\ref{finet}) at $\nu=3/2$ and the exact Read-Rezayi states for $\phi=1/3$ and $\phi=1/4$.
Lattice sizes and interaction parameters are the same as in Figs.~\ref{bosonfinetune}(c) and \ref{bosonfinetune}(d). N/D: There is no clear fourfold quasidegeneracy for this sample. \label{overlapRR2}}
\begin{ruledtabular}
\begin{tabular}{cccc}
&$N_b=9$&$N_b=12$&$N_b=15$\\
\hline
$\phi=1/3$&N/D&N/D&$0.7330$\\
$\phi=1/4$&$0.8483$&$0.6598$&$0.8849$
\end{tabular}
\end{ruledtabular}
\end{table}

\section{Candidate Systems}\label{candidate}

Having demonstrated the existence of non-Abelian FCIs in the LLL of the Kapit-Mueller Hamiltonian, we now discuss candidate physical systems in which this model could be realized. Perhaps the most natural system in which to observe the non-Abelian states that we describe here is a gas of ultracold atoms or molecules.\cite{bloch} In such experiments a dilute gas of bosons or fermions is cooled to quantum degeneracy in a magnetic or optical trap, and additional features, such as optical lattices and artificial gauge fields, are turned on adiabatically during or after the cooling process. While the atoms or molecules are themselves electrically neutral, a number of methods have been studied in recent years for generating artificial magnetic fields in them, ranging from rotation \cite{cooper,viefers} to more exotic methods involving additional lasers and multiphoton Raman transitions.\cite{hafezi,lin,spielman,liu,cooperdalibard,blochstaggered,dalibardRMP,aidelsburgeratala,Miyake} Many of these methods exploit the periodicity of an optical lattice to reach high effective flux densities and, so, are fully compatible with the Hamiltonian studied in this work.

As the trapped atoms are electrically neutral, their natural interactions are weak and short ranged, though the interaction strengths can be increased by increasing the depth of the optical lattice or by tuning the external magnetic field to exploit a Feshbach resonance. Longer-ranged interactions can be achieved by trapping ultracold polar molecules, which have a strong dipole-dipole interaction that we have shown is particularly useful in stabilizing non-Abelian ground states. Recent experiments have cooled ultracold molecules below quantum degeneracy and trapped them in 2D layers \cite{demiranda} and in optical lattices.\cite{chotia} Artificial gauge fields in polar molecules have yet to be demonstrated, but are certainly possible and are an active area of theoretical research.\cite{yaolaumann,yaodipo,manmana}

We note also that an alternative method for obtaining a lattice LLL of bosons is to engineer arrays of qubits or nonlinear optical resonators \cite{hafezidemler,nunnenkamp,kochhouck,umucalilarcarusotto,hafezilukin,kapitgauge} with specially tuned interactions and couplings to create an artificial gauge field. However, it is important to understand that the results of our bosonic calculations in this work are not directly applicable to the qubit array case. In this study we have projected Hamiltonian (\ref{HKM}) onto its lowest band and diagonalized the interaction term in this reduced subspace, a technique which implicitly assumes that the interaction energy is small compared to the band gap. In qubit arrays, the on-site interaction between bosons is typically hard core (and thus infinite energy), and while finite-ranged interactions such as potential terms between NNs or NNNs can be weak and tunable, the hard-core interaction will lead to significant ground-state mixing with the excited bands for any $\nu > 1/2$. While we expect that the interactions and flux density in qubit arrays could be tuned to achieve the same sequence of topological states which we have demonstrated here for weaker interactions, additional numerical calculations which take the hard-core constraint into account are needed to verify this claim.

\section{Conclusion}\label{con}

We have demonstrated that a realistic physical Hamiltonian for interacting lattice bosons in a uniform magnetic field can exhibit non-Abelian anyon ground states. We have shown that the MR state at $\nu = 1$ can be observed robustly for a variety of flux densities and that the more exotic RR state at $\nu = 3/2$ can also be observed at slightly lower flux ranges. The observation of the higher order RR state is a particularly exciting result, since unlike the MR state, its excitations are capable of universal topological quantum computation; however, our evidence for its existence is not as conclusive as it is for the MR state. Nonetheless, our results clearly demonstrate that non-Abelian FCIs are realistic ground-state configurations of bosons with long-range two-body interactions, further strengthening the analogy between Chern bands in the lattice and the LLL of the continuum. Our results also demonstrate the utility of using approximate Haldane pseudopotentials to guide the tuning of two-body interactions to stabilize non-Abelian ground states. In cold atoms both in optical lattices and in qubit arrays such tuning is possible by externally adjusting the physical parameters of the system, and could be vital for successfully realizing such states in real experiments.

Finally, we note that we only studied systems with a single species of boson in this work, and the addition of a spin or flavor degree of freedom unlocks new ground states. In particular, for two species of bosons with a local (flavor-independent) three-body repulsive interactions the ground state at $\nu = 4/3$ is a ``non-Abelian spin singlet'' state \cite{Ardonne2,ardonneschoutens} with Fibonacci anyon excitations. We expect that this state, as well as other interesting configurations, could also be stabilized by tuning ranged two-body interactions in the Kapit-Mueller Hamiltonian. Similarly, long-range interactions in flat band models with higher Chern numbers, $|C|>1$, are likely to result in new phases possibly including the intriguing non-Abelian states \cite{highCnonAbelian} that were recently shown to be stabilized by multi-body interactions in the pyrochlore slab (kagome multilayer) model.\cite{Trescher} We also very recently learned of a new Chern band model by Ataki\c{s}i and Oktel,\cite{atakisioktel} which, like the Kapit-Mueller Hamiltonian, obtains an exact LLL in the lattice and is also capable of flattening the excited bands. Since the ground-state bands of the two Hamiltonians are identical, our calculation should apply directly to that model, and it would also be interesting to apply our prescriptions for interaction flattening to look for correlated states in higher flat Chern bands.

\acknowledgments
Z.~L. and E.~J.~B.~thank Dmitry Kovrizhin, Andreas L\"auchli and Roderich Moessner for useful discussions during related collaborations. Z.~L.~was supported by the Department of Energy, Office of Basic Energy Sciences, through Grant No.~DE-SC0002140 and by China Postdoctoral Science Foundation Grant No.~2012M520149 and would like to thank Hong-Gang Luo at Lanzhou Unversity for the computational resources. E.~K.~was supported by EPSRC Grant Nos. EP/I032487/1 and EP/I031014/1 and by Oxford University, and would like to thank Kaden Hazzard for useful discussions. E.~J.~B.~was supported by the Emmy Noether program (BE 5233/1-1) of the German Research Foundation (DFG).

\bibliography{EDproject}
\end{document}